\documentstyle[aps,prd,twocolumn,axodraw,epsfig]{revtex}

\def\beq{\begin{equation}}
\def\eeq#1{\label{#1}\end{equation}}
\def\eeqn{\end{equation}}
\def\beqa{\begin{eqnarray}}
\def\eeqa#1{\label{#1}\end{eqnarray}}
\def\eeqan{\end{eqnarray}}
\def\CR{\nonumber \\ }
\def\leqn#1{(\ref{#1})}

\def\to{\rightarrow}
\def\eps{\epsilon}
\def\Tr{{\rm Tr}}
\newcommand\iden{\leavevmode\hbox{\small1\normalsize\kern-.33em1}}
\def\psip{\psi^\prime}
\def\sqt#1{\frac{#1}{\sqrt{2}}}
\def\tw{\theta_{\rm W}}
\def\Mz{M_{\rm Z}}
\def\Mw{M_{\rm W}}
\def\W3{W_H^3}
\def\l{\ell}

\begin{document}

\wideabs{
\begin{flushright}
LBNL-51881 \\
SLAC-PUB-9611 \\
{\tt hep-ph/0212228} \\
\end{flushright}

\title{Collider Tests of the Little Higgs Model}
\author{Gustavo Burdman$^1$, Maxim Perelstein$^1$, and Aaron Pierce$^2$.}
\address{
$^1$Theory Group, 
Lawrence Berkeley National Laboratory, 
Berkeley, CA~94720 \\
$^2$ Theoretical Physics Group, 
Stanford Linear Accelerator Center, Stanford, CA~94309.
}
\date{\today}
\maketitle

\begin{abstract}
The little Higgs model provides an alternative to traditional candidates 
for new physics at the TeV scale. The new heavy gauge bosons 
predicted by this model should be observable at the Large Hadron Collider 
(LHC). We discuss how the LHC experiments could test the little Higgs model 
by studying the production and decay of these particles.      

\end{abstract}
}

{\it Introduction ---} The hierarchy problem, the question of the origin and 
radiative stability of 
the enormous hierarchy between the electroweak symmetry breaking scale and
the fundamental scale of gravity, requires new physical phenomena to
appear at energy scales around TeV. Several models of new physics at the TeV 
scale have been proposed over the years, relying on supersymmetry, new
strong dynamics, or extra dimensions to resolve the hierarchy problem. The
experiments at the upcoming Large Hadron Collider (LHC) are expected to
determine which of these possibilities is realized in Nature.

All models of TeV-scale physics are strongly constrained by non-observation
of virtual effects of new particles in experiments at presently
available energies. For example, supersymmetric theories require non-trivial 
model building to satisfy constraints from flavor-changing neutral currents 
(FCNC). Recently, an elegant and general way to avoid such indirect 
constraints has been suggested~\cite{LH1,LH2}. In this proposal, the 
hierarchy problem is solved in two steps. First, the Standard Model (SM) in incorporated
into a ``little Higgs'' theory that possesses enough symmetry to cancel the 
quadratically divergent contribution to the Higgs mass parameter at {\it 
one-loop order.} This cancellation allows the little Higgs theory to be 
valid up to a scale of order 10 TeV without any fine tuning. Around 10 TeV,
the little Higgs theory itself breaks down, and the full dynamics resolving
the hierarchy problem becomes apparent. This dynamics may be described by 
one of the traditional solutions of the hierarchy problem, e.g. a
supersymmetric theory. The simple structure of the little Higgs theory, 
combined with the high energy scale at which the full set of new particles 
and interactions become relevant, implies that this scenario can easily 
satisfy indirect constraints such as FCNC.

\if
The little Higgs theories also provide an attractive mechanism of electroweak 
symmetry breaking (EWSB). The symmetries of the theory enforce the vanishing 
of the Higgs mass-squared parameter at tree level. The one-loop 
renormalization of this parameter is dominated by the {\it negative},
logarithmically divergent contribution from fermion loops, triggering EWSB.
\fi

The little Higgs models necessarily contain new particles whose virtual 
contributions cancel the one-loop quadratic divergence in the Higgs mass 
parameter. To avoid fine tuning, these particles should not be much 
heavier than 1 TeV; thus, they are expected to be observable at the LHC.
Crucially, the couplings of these particles to the SM Higgs are firmly
predicted: indeed, the cancellation of quadratic divergences can only 
occur for particular values of these couplings. In this letter, we will
discuss how the LHC experiments can not only discover the new states 
predicted by the little Higgs models, but also verify the structure of        
the model by measuring their couplings.

{\it The Model ---} We will concentrate on the so-called
``littlest Higgs'' model~\cite{LH1}, based on a non-linear $\sigma$
model describing $SU(5)\to SO(5)$ symmetry breaking. The symmetry-breaking
vacuum expectation value (vev) is proportional to 
\beq
\Sigma_0 = \left(\begin{array}{ccc}
0 & 0 & \iden\\                                    
0 & 1 & 0\\                                
\iden & 0 & 0 \end{array} 
\right),
\eeq{sigma0}
where $\iden$ is a 2$\times$2 identity matrix. The low-energy dynamics is
described in terms of the non-linear sigma model field
\beq
\Sigma(x) = e^{i\Pi/f}\Sigma_0 e^{i\Pi^T/f} = e^{2i\Pi/f} \Sigma_0,
\eeq{sigma}
where $f \sim$ 1 TeV is the decay constant, and $\Pi = \sum_a \pi^a(x) X^a$. 
The sum runs over the 14 broken $SU(5)$ 
generators $X^a$, and $\pi^a(x)$ are the Goldstone bosons. Furthermore, the
$[SU(2)\times U(1)]^2$ subgroup of the $SU(5)$ symmetry is gauged; the
gauged generators are given by
\beq
Q_1^a = \left( \begin{array}{ccc} \sigma^a/2& & \\ & & \\ & & \end{array}
\right),~~~~Q_2^a = \left( \begin{array}{ccc} & & \\ & & \\ & &-\sigma^{a*}/2 
\end{array} \right) 
\eeq{gauged}    
for the $SU(2)$ factors, and $Y_1 =$diag$(-3,-3,2,2,2)/10$, $Y_2=$diag
$(-2,-2,-2,3,3)/10$ for the $U(1)$'s. The gauge kinetic term has the  
form
\beq
\frac{f^2}{8} \Tr (D_{\mu} \Sigma) (D^{\mu} \Sigma)^\dagger.
\eeq{kinterm}
The covariant derivative of the $\Sigma$ field is given by
\beqa
D_{\mu} \Sigma = \partial_{\mu} \Sigma &-& i\sum_{j=1}^2 g_{j} W^{a}_{j}
(Q^{a}_{j} \Sigma + \Sigma Q^{a T}_{j}) - \CR & &
i \sum_{j=1}^2 g_{j}^{\prime} B_{j}(Y_{j} \Sigma + 
\Sigma Y_{j}), 
\eeqa{cov_der}
where $W^a_j$ ($a=1\ldots3$) and $B_j$ are the $SU(2)$ and $U(1)$ gauge 
fields, respectively,
and $g_j$ and $g^\prime_j$ are the corresponding gauge couplings.

The vev \leqn{sigma0} breaks the gauge group down to the diagonal 
$SU(2)\times U(1)$, identified with the SM electroweak group. The linear 
combinations of the gauge bosons that remain massless are given by
$W_L^a = \sin\psi\,W_1^a + \cos\psi\,W_2^a$ and $B_L = \sin\psip\,B_1 + 
\cos\psip\,B_2$, where the mixing angles are determined by
\beq
\tan\psi = g_2/g_1,~~~\tan\psip = g_2^\prime/g_1^\prime.
\eeq{mixings}
The SM $SU(2)_L$ and $U(1)_Y$ gauge couplings are given 
by $g=g_1\sin\psi$ and $g^\prime=g_1^\prime\sin\psip$, respectively.
The orthogonal linear combinations of the gauge bosons, $W_H^a$ and $B_H$, 
acquire masses 
\beq
M(W_H^a) = \frac{g}{\sin 2\psi}\,f,~~~~~M(B_H) = \frac{g^\prime}{\sqrt{5}
\sin 2\psip}\, f. 
\eeq{masses}

Out of the 14 Goldstone bosons of the $SU(5)/SO(5)$ sigma model, four are 
eaten by $W_H^a$ and $B_H$; the Goldstone boson matrix for the remaining 10
fields has the form
\beq
\Pi = \left( \begin{array}{ccc} & \sqt{H^\dagger}& \phi^\dagger\\ 
                           \sqt{H}&                & \sqt{H^*}   \\ 
                             \phi & \sqt{H^T}      &     \end{array}
\right), 
\eeq{higgses}    
where $H = (h^+, h_0)$ is the SM Higgs doublet, and $\phi$ is an additional 
Higgs field transforming as a triplet under the SM $SU(2)_L$. At tree level,
both $H$ and $\phi$ are massless. The couplings of the $H$ field to 
the gauge bosons follow from \leqn{kinterm}, and at leading order in $1/f$ 
are given by
\beq
\frac{1}{4}\,H\,\left( g_1 g_2 \,W_1^{\mu a} \,W_{2\mu}^a + 
g_1^\prime g_2^\prime \,B_1^\mu \,B_{2\mu}\right)\, H^\dagger.
\eeq{higgs_coupl}
In a generic theory, one would also expect diagonal couplings such as
$g_1^2 W_1^2 HH^\dagger$ and $g_2^2 W_2^2 HH^\dagger$. These couplings 
generally lead to quadratically divergent renormalization of the Higgs mass 
parameter. The structure of Eq.~\leqn{higgs_coupl}, in which the diagonal 
couplings are absent, guarantees the absence of such divergences, and is 
therefore a crucial feature of the little Higgs model. In terms of the  
gauge boson mass eigenstates, Eq.~\leqn{higgs_coupl} reads,
\beqa 
& &\frac{1}{4}\,H\,\bigl(g^2(W_{L\mu}^aW^{\mu a}_L - W_{H\mu}^aW^{\mu a}_H
- 2\cot 2\psi\, W_{H\mu}^aW^{\mu a}_L)
+ \CR & &~~~g^{\prime 2}(B_{L\mu}B^\mu_L - B_{H\mu}B^\mu_H
- 2\cot 2\psip\,B_{H\mu}B^\mu_L)\bigr)\,H^\dagger. 
\eeqa{higgs_coupl1}
In this basis, the diagonal couplings of the Higgs to the gauge bosons are
present, but the couplings to the light and heavy states have equal magnitude
and opposite sign, ensuring the cancellation of quadratic divergences. 
There is no such cancellation for the triplet $\phi$, which acquires a mass 
of order TeV.

Spontaneous electroweak symmetry breaking (EWSB) arises due to 
the logarithmically divergent negative contribution to the Higgs mass-squared 
from top loops~\cite{LH1}. As usual, we parametrize $H = U(x)\,\sqt{1}
(0, v+h(x))$, and choose the unitary gauge, $U(x)=1$. The ratio of 
the Higgs vev and the decay constant $f$ provides a useful expansion 
parameter: $\eps=v/f \lesssim 0.2$. The triplet $\phi$ is generally also 
expected 
to acquire a small vev, $v^\prime\ll v$~\cite{EWP1}; we will neglect this 
effect in our analysis. 

{\it Higgs-gauge boson couplings ---} EWSB induces cubic couplings between the
physical Higgs boson and the gauge bosons of the model. The couplings can be
obtained from Eq.~\leqn{kinterm}. Consider first the neutral gauge bosons.
Let $V=(W_L^3, B_L, W_H^3, B_H)$; the relevant terms in \leqn{kinterm} can
be written as
\beq
\frac{1}{2}\,f^2 V^\mu \,\left({\cal M}_1 + \,\left(\eps+\frac{h(x)}{f}
\right)^2 {\cal M}_2 + \ldots \right)\,V^T_\mu,
\eeq{neutral}
where ${\cal M}_1=$diag$(0,0,g^2/\sin^22\psi, g^{\prime2}/5\sin^22\psip)$,
and the dots represent terms of higher order in $1/f$. The gauge boson
mass matrix to second order in $\eps$, ${\cal M}_1+\eps^2 {\cal M}_2$, is 
not diagonal in the chosen basis; to diagonalize it, we perform the 
standard rotation by the Weinberg angle $\tw$ in the upper left corner 
taking $(W_L^3, B_L)\to(A, Z)$. All 
\vskip120pt
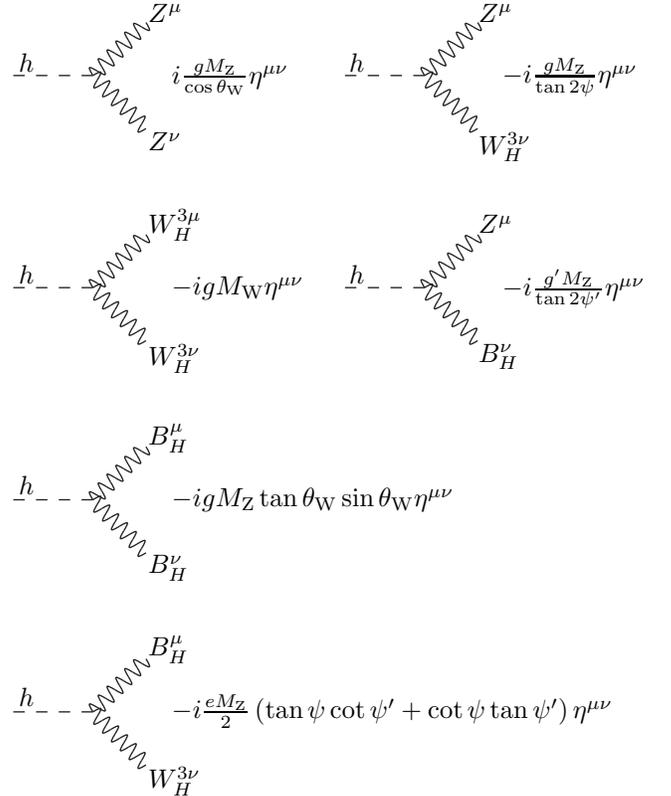
\begin{figure}[b]
\begin{center} 
\begin{picture}(230,200)(0,-120)
  \Text(2,162)[bl]{$h$}
  \DashLine(0,160)(30,160){4}
  \Text(51,181)[bl]{$Z^\mu$}
  \Photon(30,160)(50,180){3}{6}
  \Text(51,139)[tl]{$Z^\nu$}
  \Photon(30,160)(50,140){3}{6}
  \Text(60,160)[l]{$i {g\Mz \over \cos\tw} \eta^{\mu\nu}$}
  \Text(127,162)[bl]{$h$}
  \DashLine(125,160)(155,160){4}
  \Text(176,181)[bl]{$Z^\mu$}
  \Photon(155,160)(175,180){3}{6}
  \Text(176,139)[tl]{$W_H^{3\nu}$}
  \Photon(155,160)(175,140){3}{6}
  \Text(185,160)[l]{$-i {g\Mz \over \tan 2\psi} \eta^{\mu\nu}$}

  \Text(2,82)[bl]{$h$}
  \DashLine(0,80)(30,80){4}
  \Text(51,101)[bl]{$W_H^{3\mu}$}
  \Photon(30,80)(50,100){3}{6}
  \Text(51,59)[tl]{$W_H^{3\nu}$}
  \Photon(30,80)(50,60){3}{6}
  \Text(60,80)[l]{$-i g\Mw \eta^{\mu\nu}$}
  \Text(127,82)[bl]{$h$} \DashLine(125,80)(155,80){4}
  \Text(176,101)[bl]{$Z^\mu$}
  \Photon(155,80)(175,100){3}{6}
  \Text(176,59)[tl]{$B_H^\nu$}
  \Photon(155,80)(175,60){3}{6}
  \Text(185,80)[l]{$-i {g^\prime \Mz \over \tan 2\psip} \eta^{\mu\nu}$}

  \Text(2,2)[bl]{$h$}
  \DashLine(0,0)(30,0){4}
  \Text(51,21)[bl]{$B_H^\mu$}
  \Photon(30,0)(50,20){3}{6}
  \Text(51,-21)[tl]{$B_H^\nu$}
  \Photon(30,0)(50,-20){3}{6}
  \Text(60,0)[l]{$-i g\Mz \tan\tw \sin\tw \eta^{\mu\nu}$}

  \Text(2,-78)[bl]{$h$}
  \DashLine(0,-80)(30,-80){4}
  \Text(51,-59)[bl]{$B_H^\mu$}
  \Photon(30,-80)(50,-60){3}{6}
  \Text(51,-101)[tl]{$W_H^{3\nu}$}
  \Photon(30,-80)(50,-100){3}{6}
  \Text(60,-80)[l]{$-i \frac{e\Mz}{2} \left(\tan\psi \cot\psip + 
\cot\psi \tan\psip \right) \eta^{\mu\nu}$}
\end{picture}
\caption{Cubic couplings between the physical Higgs boson $h$ and pairs of 
neutral gauge bosons.}
\label{fig:1}
\end{center}
\end{figure}
\newpage
\vskip-40pt
\begin{figure}[b]
\begin{center} 
\begin{picture}(230,120)(0,-40)
  \Text(2,82)[bl]{$h$}
  \DashLine(0,80)(30,80){4}
  \Text(51,101)[bl]{$W_L^{+\mu}$}
  \Photon(30,80)(50,100){3}{6}
  \Text(51,59)[tl]{$W_L^{-\nu}$}
  \Photon(30,80)(50,60){3}{6}
  \Text(60,80)[l]{$i g\Mw \eta^{\mu\nu}$}
  \Text(127,82)[bl]{$h$}
  \DashLine(125,80)(155,80){4}
  \Text(176,101)[bl]{$W_H^{+\mu}$}
  \Photon(155,80)(175,100){3}{6}
  \Text(176,59)[tl]{$W_H^{-\nu}$}
  \Photon(155,80)(175,60){3}{6}
  \Text(185,80)[l]{$-i g \Mw \eta^{\mu\nu}$}

  \Text(2,2)[bl]{$h$}
  \DashLine(0,0)(30,0){4}
  \Text(51,21)[bl]{$W_H^{\pm\mu}$}
  \Photon(30,0)(50,20){3}{6}
  \Text(51,-21)[tl]{$W_L^{\mp\nu}$}
  \Photon(30,0)(50,-20){3}{6}
  \Text(60,0)[l]{$-i g\Mw \cot 2\psi\,\eta^{\mu\nu}$}
\end{picture}
\caption{Cubic couplings between the physical Higgs boson $h$ and pairs of 
charged gauge bosons.}
\label{fig:2}
\end{center}
\end{figure}
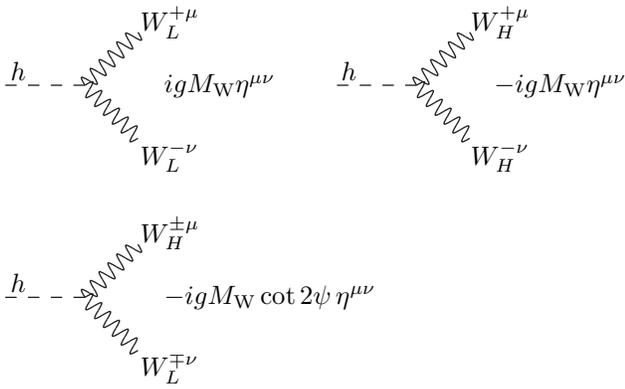
\noindent other rotations required to diagonalize the mass matrix involve 
angles
of order $\eps$; we will not need to know them explicitly. The Higgs 
couplings to pairs of gauge bosons are proportional to the elements of the
matrix ${\cal M}_2$; we collect the non-vanishing couplings in Fig.~1,
keeping only the leading order in $\eps$. 
Note that ${\cal M}_2$ is {\it not} diagonal in the mass eigenbasis, leading 
to off-diagonal Higgs couplings such as $\W3 Z h$. 
The cubic couplings of the Higgs to charged gauge bosons are computed in the
same fashion; they are shown in Fig.~2. 

The three diagonal couplings of the Higgs in Fig.~1, $hZZ$, $h\W3 \W3$ and 
$hB_HB_H$, add up to zero. So do the two diagonal couplings in Fig.~2, 
$hW_L^+W_L^-$ and $hW_H^+W_H^-$. These cancellations can be traced back to 
Eq.~\leqn{higgs_coupl1}, and are therefore directly related to
the crucial cancellation of quadratic divergences. Measuring the diagonal 
couplings would provide the most direct way to verify the little Higgs model. 
Unfortunately, experimentally this is a difficult task: the measurement 
requires associated production of a $W_H$ boson with a Higgs. The cross
section for this process is very small: with 100 fb$^{-1}$ integrated
luminosity at the LHC there are typically only tens of events in this 
channel, rendering it
virtually unobservable once specific final states are considered. 
It is much easier to measure the off-diagonal couplings, such as $h \W3 Z$
and $h W_H^\pm W_L^\mp$. Although these couplings do not directly participate 
in the cancellation of quadratic divergences, verifying their structure
would provide a strong evidence for the crucial feature of the model, 
Eq.~\leqn{higgs_coupl}. Indeed, the factor of $\cot 2\psi$ in these 
couplings is a unique consequence of Eq.~\leqn{higgs_coupl}. 
To illustrate this point, consider an alternative ``big Higgs''
theory with the same $[SU(2)\times U(1)]^2$ gauge structure as the little 
Higgs, but with a Higgs field transforming only under a single $SU(2)\times 
U(1)$ factor with SM quantum numbers. (The one-loop quadratic divergence in 
the Higgs mass parameter
is {\it not} canceled in this theory.) This theory predicts $h \W3 Z$
and $h W_H^\pm W_L^\mp$ vertices of the same form as in the little Higgs 
model, but with the replacement $\cot 2\psi \to \cot\psi$. Thus, if the
mixing angle $\psi$ can be obtained independently, the measurement of these 
vertices would act as a discriminator between the little Higgs and the 
alternative theory.      

{\it Production and decay of heavy gauge bosons ---} Heavy gauge bosons 
$W^a_H$ and $B_H$ are produced at the LHC predominantly through their coupling 
to quarks. The charges of the SM fermions under the extended $[SU(2)\times 
U(1)]^2$ gauge group are constrained by the requirement that they have 
correct transformation properties under its low-energy subgroup. We choose
the left-handed fermions to transform as doublets under $SU(2)_1$ and  
singlets under $SU(2)_2$; the right-handed fermions are singlets under both
$SU(2)$'s~\cite{another}.
The couplings of the heavy $SU(2)$ bosons have the form
\beq
g \cot\psi\,W^a_{H\mu}\,\left( \bar{L} \gamma^\mu \frac{\sigma^a}{2}\,L\,+\,
\bar{Q} \gamma^\mu \frac{\sigma^a}{2}\,Q\right),
\eeq{fermion_c}
where $L$ and $Q$ are the left-handed lepton and quark fields, and we 
suppress the generation indices.
The charges of the fermions under the two $U(1)$ groups have to add up to
the SM hypercharge, $Q_1+Q_2=Y$, but are otherwise unconstrained. This 
implies a high degree of model-dependence in the couplings of the $B_H$ 
boson to fermions.  Note that the strongest electroweak precision 
constraints on the little Higgs model~\cite{EWP1,EWP2} arise precisely
from the 
diagrams involving the $B_{H}$. Certain choices of the fermion charges 
and the angle $\psip$ can help minimize these constraints~\cite{EWP3}; 
alternately, the extra $U(1)$ can be eliminated completely~\cite{pc} without 
introducing significant fine tuning due to the smallness of the coupling
$g^{\prime}$. Given this model uncertainty in the $U(1)$ sector, we will 
concentrate on the production and decay of the $SU(2)$ heavy bosons.

\begin{figure}
\begin{center}
\includegraphics[scale=0.35,angle=90]{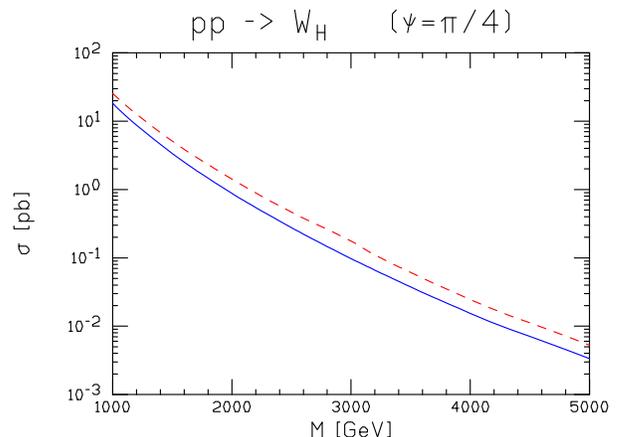}%
%
\caption{Production cross sections for $\W3$ (solid) and $W_H^\pm$ (dashed) 
at the LHC, for $\psi=\pi/4$.  We use the CTEQ5L parton distribution function.
}
\end{center}
\label{fig1}
\end{figure}

In $pp$ collisions at the LHC, the heavy gauge bosons are predominantly 
produced through $q \bar{q}$ annihilation. The sub-process $q g \to W_H q$
is considerably smaller and could be separately identified due to the presence
of a high $p_T$ jet. Fig.~3 shows the leading order production cross section 
of $W_H$ as a function of its mass, for the case $\psi=\pi/4$. 
The general case may be obtained from Fig.~\ref{fig1} by simply 
scaling by $\cot^2\psi$. 
From our discussion thus far, it is clear that the two-body 
decay channels of the $\W3$ include 
$\W3\to \bar{f}f$, where $f$ is any of the SM quarks or leptons, as well
as $\W3\to Zh$. (The decay $\W3\to B_H h$ may also be allowed
in certain regions of the parameter space.) Partial decay widths in 
these channels are easily computed using the Feynman rules in Fig.~1:
\beqa
\Gamma(\W3\to \bar{\l}\l) &=& \frac{g^2 \cot^2\psi}{96\pi}\,M, \CR
\Gamma(\W3\to \bar{q}q) &=& \frac{g^2 \cot^2\psi}{32\pi}\,M, \CR
\Gamma(\W3\to Zh) &=& \frac{g^2 \cot^2 2\psi}{192\pi}\,M, 
\eeqa{widths}
where $M=gf/\sin2\psi$ is the mass of the $W_H$ triplet, and we neglect 
corrections of order $\eps=v/f$ (including the effects of non-zero top mass). 

In addition, the decay mode $W^{3}_H \to W^{+}_{L} W^{-}_{L}$ is also 
important~\cite{Logan}. In the unitary gauge, the triple gauge boson 
vertex $W^{3}_H W^{+}_{L} W^{-}_{L}$ arises from expressing the gauge field 
strengths of the two $SU(2)$ groups in terms of mass eigenstates. Although this 
vertex is of order $(v/f)^{2}$, this suppression is compensated 
by enhancements of order $f/v$ coming from the longitudinal polarization 
vectors of the $W$ bosons.\footnote{Similar enhancement also occurs for the 
inverse process, creation of a $W_{3}$ by $W$-fusion. However, this process
is still completely subdominant to the Drell-Yan production of $W_3$'s due to 
smallness of $W$ luminosity at high $x$.}  The decay width for this mode 
is given by
\beq
\Gamma(\W3\to W^{+}_{L} W^{-}_{L}) = \frac{g^2 \cot^2 2\psi}{192\pi}\,M.
\eeq{W3WW} 
Note that this is equal to the width for the $Zh$ decay mode, as expected 
from the Goldstone boson equivalence theorem.  

Summing over all the quark and lepton channels and ignoring the $B_H h$
mode (and all $N$-particle final states with $N\geq3$) results in a total 
width
\beq
\Gamma_{\rm tot} = \frac{g^2}{96\pi}\,\left(\cot^2 2\psi + 24 \cot^2\psi
\right)\,M. 
\eeq{total_w}
Partial decay widths of the $W_H^{\pm}$ bosons are easily obtained 
from~\leqn{widths} using the isospin symmetry, which is accurate to leading 
order in $\eps$.

{\it Testing the model at the LHC ---} The discovery reach of the LHC for
the $\W3$ and $W^{\pm}_{H}$ gauge bosons is quite high. 
The cleanest mode is $\W3\to\ell^+\ell^-$, with $\ell=e$~or~$\mu$.
Existing $Z^\prime$ studies at the LHC~\cite{atlas}
indicate that these channels are virtually free of backgrounds. 
If we establish the discovery reach requiring the observation of $10$~events,
than for 100~fb$^{-1}$ luminosity this translates into a mass reach of 
approximately $M=5\cdot(\cot \psi)^{1/3}~$TeV for $\psi=\pi/4$. 
The lower bound on $M$ in a model with only a single gauged $U(1)$ from 
electroweak precision constraints is roughly 2.5 TeV at 95\% C.L., and is 
approximately independent of the mixing angle.  Thus, there is a wide 
range of allowed parameters for which the extra gauge bosons are observable.   

Discovering the $W_H$ triplet does not by itself provide a striking signature 
for the little Higgs model. Indeed, one can imagine many alternative 
theories in which such a triplet is present. As we already mentioned, a
good way to distinguish between the little Higgs and alternative theories 
is to independently measure the mixing angle $\psi$ and the magnitude of
the $h\W3 Z$ and $h W_H^\pm W_L^\mp$ couplings. Let us outline how these
measurements can be performed at the LHC.

The number of events in the $pp\to\W3\to \l^+\l^-$ channel, 
where $\l=e$ or $\mu$, is given by
\beq
N(\l^+ \l^-) = {\cal L}\,\sigma_{\rm prod}^{(0)} \,f(\tan^2\psi),
\eeq{ll_num}  
where ${\cal L}$ is the integrated luminosity, $\sigma_{\rm prod}^{(0)}$ is
the $\W3$ production cross section for $\psi=\pi/4$ and fixed $M$ (see Fig. 3),
and $f(x)=4x^{-1}[96+(1-x)^2]^{-1}$. Assuming that $\sigma_{\rm prod}^{(0)}$ 
can be accurately predicted, measuring the $\l^+\l^-$ event rates yields 
$f(x)$, which can then be inverted to find the angle $\psi$ (up to a discrete 
ambiguity). If the $W_H$ triplet is not too 
heavy, the event rates are quite large, leading to small
statistical uncertainties. For example, for $M=2.5$~TeV, $\psi=\pi/6$, and
${\cal L}=100$ fb$^{-1}$, we find $N(e^+e^-)\approx 3500$, so the 
statistical uncertainty in measuring $\psi$ using this channel alone is less 
than 2\%. The number of $e^+e^-$ 
background events with such a high invariant mass is very small.
The most important limiting factor in this measurement is the
lack of precise knowledge of $\sigma_{\rm prod}^{(0)}$, 
limited by the uncertainties in parton distribution functions.
An alternative possibility is to measure the total width of the $\W3$ and 
use Eq.~\leqn{total_w} to extract $\psi$; the uncertainty in this measurement 
is dominated by finite calorimeter resolution.

The number of events in each of the 
$pp\to\W3\to \bar{q}q$ channels is given by $3N(\l^+\l^-)$. Verifying this 
relation provides a good test of the universality of the $\W3 \bar{f} f$ 
couplings, Eq.~\leqn{fermion_c}. Once the universality is confirmed, these
channels can be combined to further reduce the statistical uncertainty in the 
$\psi$ determination.  

Once the mixing angle $\psi$ is determined, one simply needs to count the 
number of events in the $pp\to\W3\to Zh$ channel. The little Higgs model
predicts  
\beq
N(Zh) = {\cal L}\,\sigma_{\rm prod}^{(0)} \,g(\tan^2\psi),
\eeq{Zh_num}  
where $g(x)=(1-x)^2 x^{-1}[192+2 (1-x)^2]^{-1}$. Taking again the sample
parameter values, $M=2.5$~TeV, $\psi=\pi/6$, and ${\cal L}=100$ fb$^{-1}$, 
we obtain $N(Zh)\approx 200$. The alternative ``big Higgs'' theory predicts 
$N(Zh) = 0.5 N(l^+ l^-)$, or about 10 times more $Zh$ events for the sample 
parameter values than in the little Higgs model. (We assume that the 
fermion couplings to the gauge bosons of the ``big Higgs'' model are identical
to Eq.~\leqn{fermion_c}.) 

In order to evaluate the observability of this 
mode, we have assumed that the Higgs boson is heavy enough to decay into a 
$W$ pair and considered the process~\cite{lightH} 
\beq
\W3 \to Zh \to Z~W^+ W^- \to 4j + e\nu.
\eeq{decay} 
We use the package MadGraph~\cite{madgraph} to estimate the corresponding 
backgrounds. We require that the kinematics of the observed final state
be consistent with the decay chain~\leqn{decay}. (Note that both the $\W3$
mass and the Higgs boson mass will be known fairly well by the time this
experiment becomes feasible.) We find that this requirement reduces the 
background to manageable levels, allowing one to clearly distinguish    
between the little Higgs model with the sample parameter values and the
``big Higgs'' alternative with 100 fb$^{-1}$ integrated luminosity. 

While in the above discussion we concentrated on the $\W3$ boson, the same
measurements can be performed using its charged partners $W_H^\pm$. Once 
the isospin structure of the couplings is verified, the measurements
in $\W3$ and $W_H^\pm$ channels can be combined to further improve the 
statistics. 

{\it Conclusions ---} If the little Higgs model is a part of the solution to 
the hierarchy problem, the extra gauge bosons present in this model should
be light enough to be copiously produced at the LHC. The model makes
definite predictions for the couplings of these gauge bosons to the light
SM Higgs boson, which can be measured at the LHC in decays such as 
$\W3\to Zh$ and $W_H^\pm\to W_L^\pm h$. This measurement tests the detailed
structure of the Higgs sector of the model, and can be used to distinguish
it from alternative theories with the same gauge group. 

{\it Acknowledgments ---} We would like to thank Nima Arkani-Hamed, 
JoAnne Hewett, Ian Hinchliffe, Heather Logan, Hitoshi Murayama, Michael Peskin, 
Tom Rizzo, Tim Stelzer, John Terning and Jay Wacker for 
useful discussions related to this work. G.~B. and M.~P. are supported 
by the Director, Office of Science, Office of High Energy and Nuclear 
Physics, of the U.~S. Department of Energy under Contract DE-AC03-76SF00098. 
A.~P. is supported by the U.~S. Department of Energy under 
Contract DE-AC03-76SF00515.


\begin{thebibliography}{99}
\bibitem{another}
Another simple and consistent $SU(2)$ charge assignment corresponds to 
interchanging $SU(2)_1\leftrightarrow SU(2)_2$; the results for this case 
can be obtained by exchanging $\psi\leftrightarrow \pi/2-\psi$ in 
Eq.~\leqn{fermion_c} and below.
\bibitem{comment1} The direct Tevatron bounds of Ref.~\cite{EWP2}
are avoided, for example, in models where only a single $U(1)$ is gauged.
\bibitem{lightH} An alternative decay chain, $\W3\to Zh \to \ell^+ \ell^- +
4j$, has a branching ratio smaller than~\leqn{decay} by about a factor of 3, 
but could also be useful, especially for relatively light $\W3$. 
For a lighter Higgs, one could use the decay chain 
$\W3\to Zh \to \ell^+ \ell^- b\bar{b}$.
\end{thebibliography}
\end{document}